\documentclass[twocolumn,showpacs,preprintnumbers,amsmath,nofootinbib,amssymb]{revtex4}
\input psfig.sty
\def \be {\begin{equation}}
\def \ee {\end{equation}}
\def \bea {\begin{eqnarray}}
\def \eea {\end{eqnarray}}

\usepackage{graphicx}% Include figure files
\usepackage{dcolumn}% Align table columns on decimal point
\usepackage{bm}% bold math

\begin{document}

\title{Cosmology with interaction in the dark sector}

\author{F. E. M. Costa}
\email{ernandes@on.br}

\author{E. M. Barboza Jr.} 
\email{edesio@on.br}

\author{J. S. Alcaniz}
\email{alcaniz@on.br}

\affiliation{Observat\'orio Nacional, 20921-400 Rio de Janeiro - RJ, Brasil}

\pacs{98.80.Cq; 95.36.+x}

\date{\today}

\begin{abstract}

Unless some unknown symmetry in Nature prevents or suppresses a non-minimal coupling in the dark sector, the dark energy field may interact with the pressureless component of dark matter. In this paper, we investigate some cosmological consequences of a general model of interacting dark matter-dark energy characterized by a dimensionless parameter $\epsilon$. We derive a coupled scalar field version for this general class of scenarios and carry out a joint statistical analysis involving SNe Ia data ({Legacy} and {Constitution} sets), measurements of baryon acoustic oscillation peak at $z = 0.20$ (2dFGRS) and $z = 0.35$ (SDSS), and measurements of the Hubble evolution $H(z)$. For the specific case of vacuum decay ($w = -1$), we find that, although physically forbidden, a transfer of energy from dark matter to dark energy is favored by the data.

\end{abstract}

\maketitle

\section{Introduction}

As is well known, there is mounting observational evidence that our Universe is presently dominated by two exotic forms of matter or energy. Cold, nonbaryonic dark matter (CDM), which accounts for $\simeq 30\%$ of the critical mass density and whose leading particle candidates are the axions and the supersymmetric charginos and neutralinos, was originally proposed to explain the general behavior of galactic rotation curves that differ significantly from the one predicted by Newtonian mechanics. Later on, it was also realized that the same concept is necessary for explaining the evolution of the observed structure in the Universe from density inhomogeneities of the size detected by a number of Cosmic Microwave Background (CMB) experiments. Dark energy or quintessence, which accounts for  $\simeq 70\%$ of the critical mass density and whose leading candidates are a cosmological constant $\Lambda$ and a relic scalar field $\Phi$, has been inferred from a combination of astronomical observations which includes distance measurements of type Ia supernovae (SNe Ia) indicating that the expansion of the Universe is speeding up, CMB anisotropy data suggesting $\Omega_T \simeq 1$, and clustering estimates providing $\Omega_m \simeq 0.3$ (see, e.g., \cite{revde} for current reviews).

Another interesting aspect related to these two dark components is that, unless some unknown symmetry in Nature prevents or suppresses a non-minimal coupling between them (see \cite{carroll} for a discussion), the dark matter and dark energy fields may interact between themselves, giving rise to the so-called models of coupled quintessence (see \cite{cq,cq1,jmaia1,saulo} and Refs. therein). From the observational side, no piece of evidence has been so far unambiguously presented against such an interaction, so that a weak coupling still below detection cannot be completely excluded. In the cosmological context, models of coupled quintessence are capable of explaining the current cosmic acceleration, as well as other recent observational results \cite{cq,jmaia1,saulo}. From the theoretical viewpoint, however, critiques to these scenarios do exist and are mainly related to the fact that in order to establish a model and study its cosmological consequences, one needs first to specify a phenomenological coupling $\Gamma$ between the cosmic components.

In this paper, instead of adopting the traditional approach, we follow the qualitative arguments used in Refs.~\cite{wm,alc,ernandes,prd} and deduce the interacting law from a simple argument about the effect of the dark energy decay on the dark matter expansion rate. As a step towards fundamental physics motivation for this phenomenological dark matter/dark energy interaction, we study how this class of coupled cosmologies can be interpreted in terms of classical scalar field dynamics.   We also discuss current observational constrains on the interacting term $\epsilon$. To this end we use two different samples of SNe Ia data, namely, the SNLS collaboration sample of 115 data points \cite{Astier06} and the most recent SNe Ia compilation, the so-called \emph{Constitution
set} of 397 SNe Ia covering a redshift range from $z = 0.015$ to $z = 1.551$ \cite{cs}. Given the complementarity of SNe Ia data with other cosmological observables, a joint analysis with measurements of baryon acoustic oscillation (\rm BAO) peak at $z = 0.20$~\cite{2df} and $z = 0.35$~\cite{bao}, and measurements of the Hubble evolution $H(z)$~\cite{svj} is also performed.

\begin{figure*}[t]
%\vspace{.2in}
\centerline{\psfig{figure=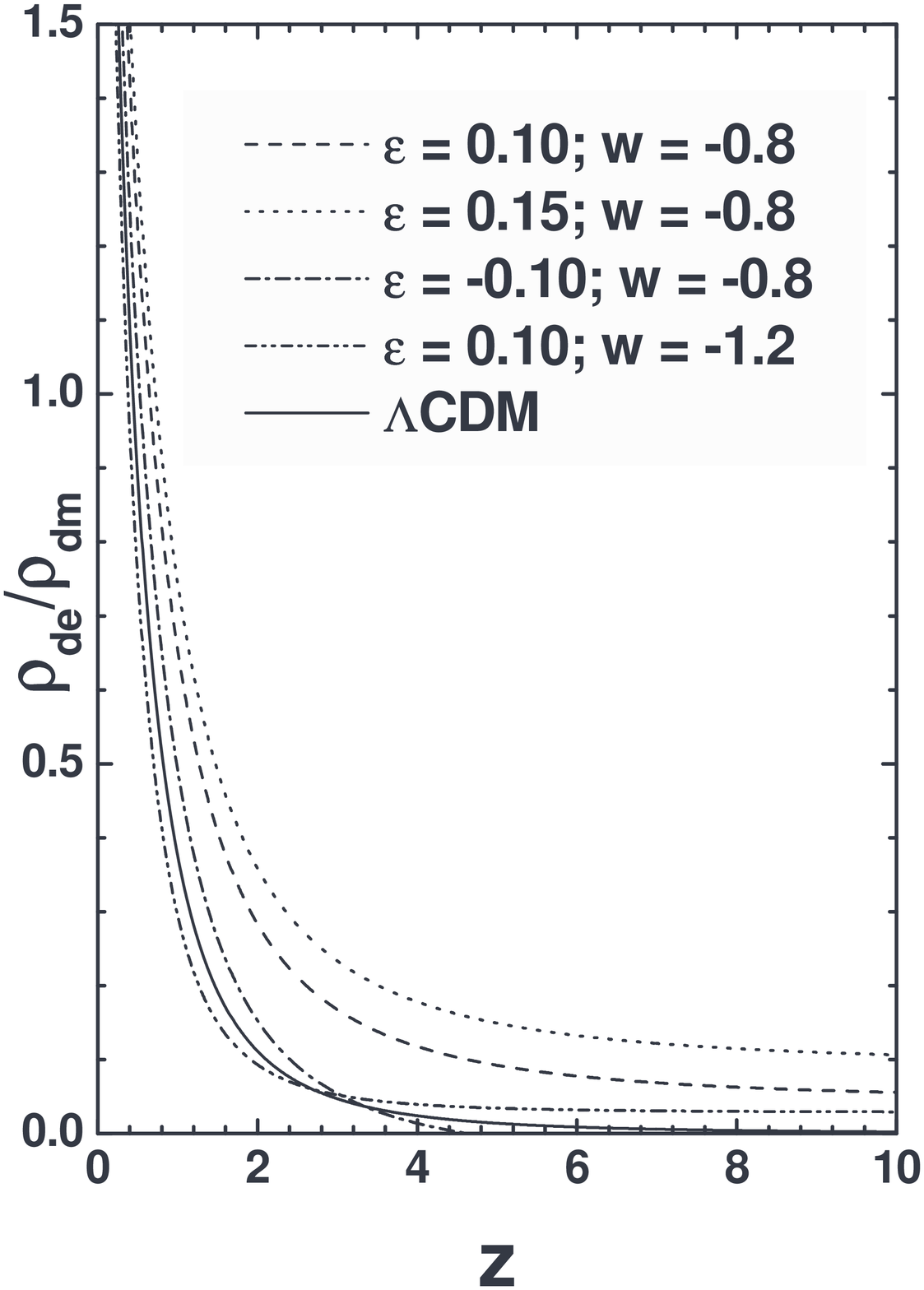,width=3.0truein,height=2.3truein}
\psfig{figure=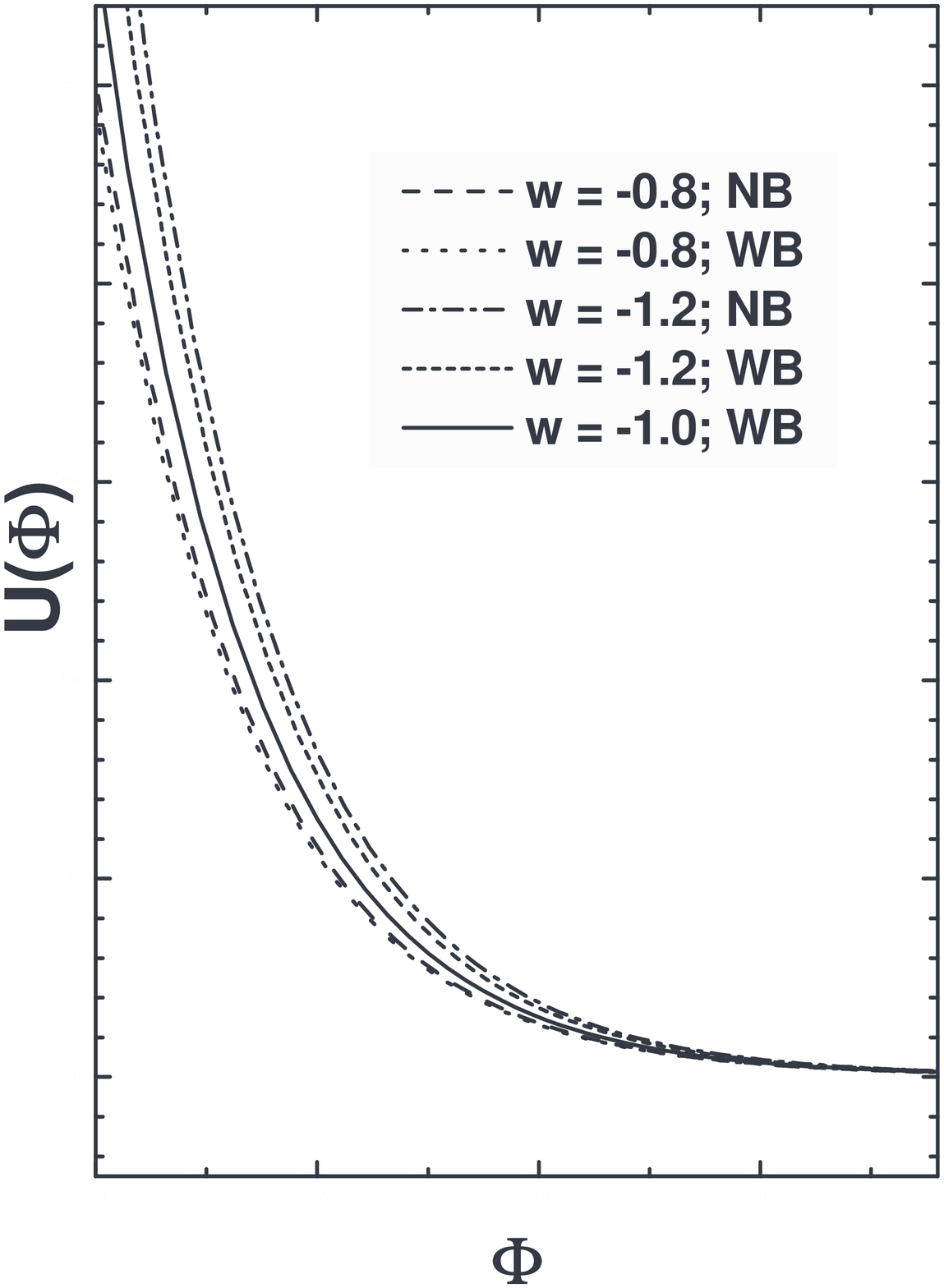,width=3.0truein,height=2.3truein}
\hskip 0.1in} 
\caption{{{\it Left:}} The ratio $\rho_{de}/\rho_{dm}$ as a function of the redshift parameter $z$ [Eqs. (\ref{dm}) and (\ref{de})] for some selected values of $\epsilon$ and $w$ and $\Omega_{de,0}/\Omega_{dm,0} \simeq 3$. Note that, for large and positive values of $\epsilon$, the relative contribution of dark energy to dark matter may be significant at early times. {{\it Right:}} The potential $U(\Phi)$ [Eqs. (\ref{ph})-(\ref{va})] as a function of the field for some selected values of $w$ and $\epsilon = 0.1$ and $\Omega_{dm}  = 0.23$. In this Panel, ``NB" means that the baryons participate of the interacting process whereas ``WB'' means that they were regarded as a separetely conserved component contributing with $\Omega_{b,0} = 0.0416$.}
\end{figure*}

\section{interacting dark matter/dark energy} 

If dark energy and dark matter interacts, the energy density of this latter component will dilute at a different rate compared to its standard evolution, $\rho_{dm} \propto a^{-3}$, where $a$ is the cosmological scale factor. Thus, the deviation from the standard dilution may be characterized by a constant $\epsilon$, such that
\begin{equation} \label{dm}
\rho_{dm} = \rho_{dm, 0}a^{-3 + \epsilon}\;,
\end{equation}
where we have set the present-day value of the cosmological scale factor $a_0 = 1$. By assuming that the radiation and baryonic fluids are separately conserved, the energy conservation law for the two interacting components reads 
%\begin{equation}
$u_{\alpha}\bar{T}^{\alpha \beta}_{;\beta} = 0$, where %, \quad \mbox{where}\quad 
$\bar{T}^{\alpha \beta} = {T}^{\alpha \beta}_{dm} + {T}^{\alpha \beta}_{de}$
%\end{equation}
or, equivalently (in our notation, $\rho_m = \rho_b + \rho_{dm}$),
\begin{equation} \label{balance}
\dot{\rho}_{dm} + 3\frac{\dot{a}}{a}\rho_{dm} = -\dot{\rho}_{de} - 3\frac{\dot{a}}{a}(\rho_{de} + p_{de})\;.
\end{equation}
By combining the above equations and considering that the dark energy component is described by an equation of state parameter $w \equiv p_{de}/\rho_{de}$, one can show that the energy density of the dark energy component evolves as \cite{prd}
\begin{equation}
\label{de}
\rho_{de} = \rho_{de, 0}a^{-3(1 + w)} + \frac{\epsilon \rho_{dm, 0}}{3|w| - \epsilon}a^{-3 + \epsilon}\;,
\end{equation}
where the integration constant $\rho_{de, 0}$ is the present-day fraction of the dark energy density. Clearly, in the absence of a coupling with the CDM component, i.e., $\epsilon = 0$, the conventional non-interacting quintessence scenario is fully recovered. For $w = -1$ and $\epsilon \neq 0$, we may identify $\rho_{de, 0} \equiv \rho_{v, 0}$ (the current value of the vacuum contribution), and the above expression reduces to the vacuum decaying scenario recently discussed in Refs.~\cite{wm,alc,ernandes}.

Fig. (1a) shows the ratio $\rho_{de}/\rho_{dm}$ as a function of the redshift parameter [Eqs. (\ref{dm}) and (\ref{de})] for some selected values of $\epsilon$ and $w$ and $\Omega_{de,0}/\Omega_{dm,0} \simeq 3$. It is shown that this fraction is a very sensitive function of the interacting parameter $\epsilon$. For instance, at $z = 5$, the dark energy contribution relative to $\rho_{dm}$ for $\epsilon = 0.15$ is about 75\% larger than the same contribution for $\epsilon = 0.10$. This amounts to saying that for relatively large and positive values of $\epsilon$ the energy density of the pressureless component is not negligible at high-$z$. For the sake of comparison, the $\Lambda$CDM case ($\epsilon = 0$ and $w = -1$) is also shown (solid line).

\section{scalar field description}

In order to find the scalar field counterpart for the flat interacting dark matter/energy model considered here (that shares the same dynamics and temperature evolution law), we use the procedure originally proposed in Ref.~\cite{jmaia} (see also \cite{jmaia1} for a general description with arbitrary curvature).

First, let us define the parameter
%\begin{equation}
$\gamma_* = - {2\dot{H}}/{3H^2} =  (1 + \omega\Omega_{de})$,
%\end{equation}
which is just another way of writing the field equations 
\begin{equation}
8\pi G(\rho_b + \rho_{dm} + \rho_{de}) = 3H^2, \label{fried}
\end{equation}
\begin{equation}
\label{friedp} 8\pi Gp_{de} =-2\dot{H}-3{H^2}\ ,
\end{equation}
where $\rho_b$, $\rho_{dm}$ and $\rho_{de}$ are the energy densities of baryons, cold dark matter and dark energy, respectively, whereas $p_{de}$ stands for the dark energy pressure. Thus, any cosmology dynamically equivalent to the  coupled quintessence scenario considered here must have the same $\gamma_*$. Following standard lines, we replace the energy density and pressure in Eqs. (\ref{fried}) and (\ref{friedp}) by the corresponding scalar field expressions, i.e.,
%\begin{equation}
$\rho_{de}  \rightarrow \rho_{\phi}$ and $p_{de}  \rightarrow p_{\phi}$,
%\end{equation}
where $\rho_{\phi} = \dot{\phi}^2/2 + V(\phi)$ and $p_{\phi} = \dot{\phi}^2/2 - V(\phi)$ are, respectively, the energy density and pressure associated with the coupled scalar field $\phi$ whose potential is $V(\phi)$.

\begin{figure*}[t]
%\vspace{.2in}
\centerline{\psfig{figure=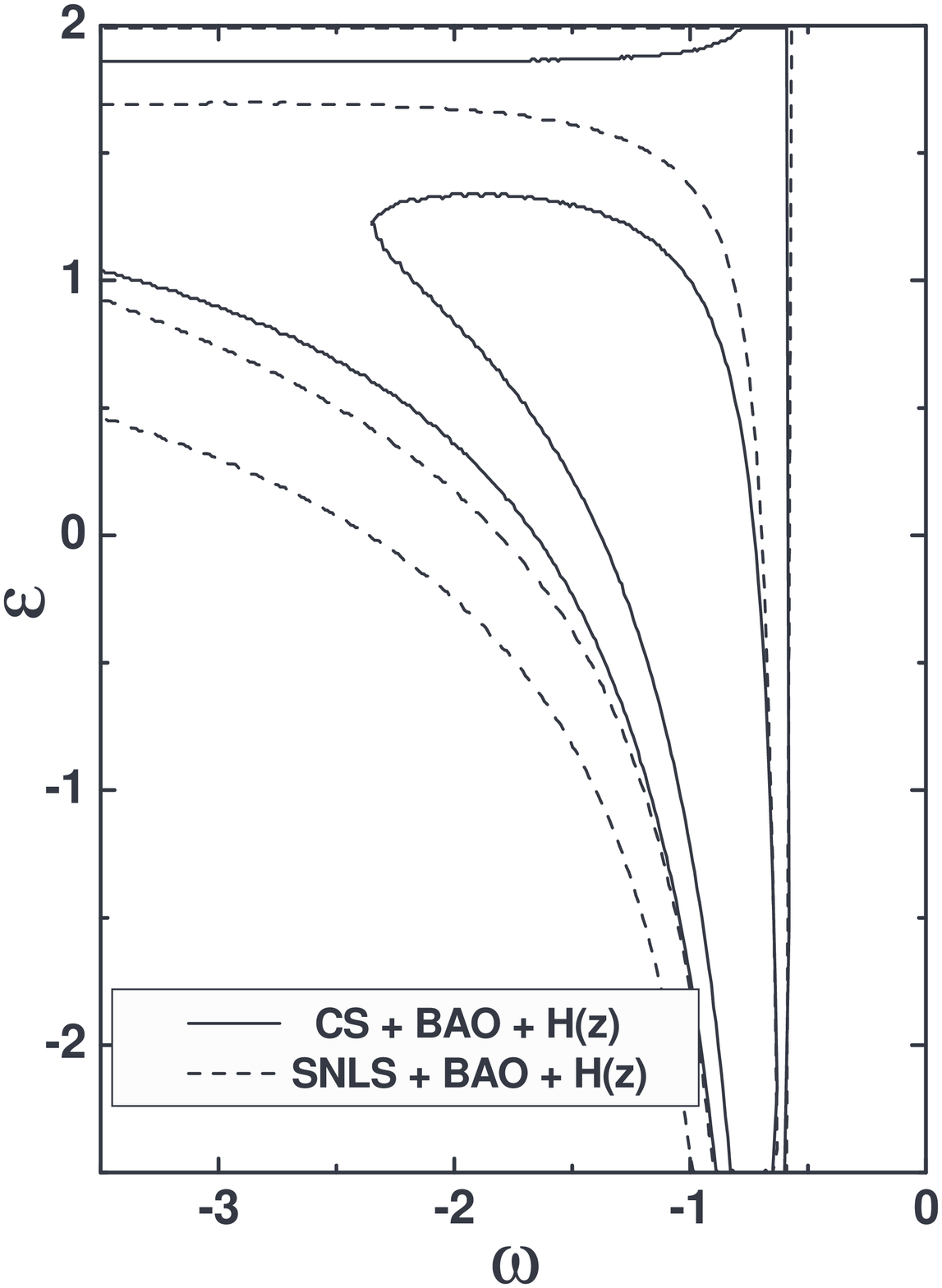,width=2.4truein,height=2.0truein}
\psfig{figure=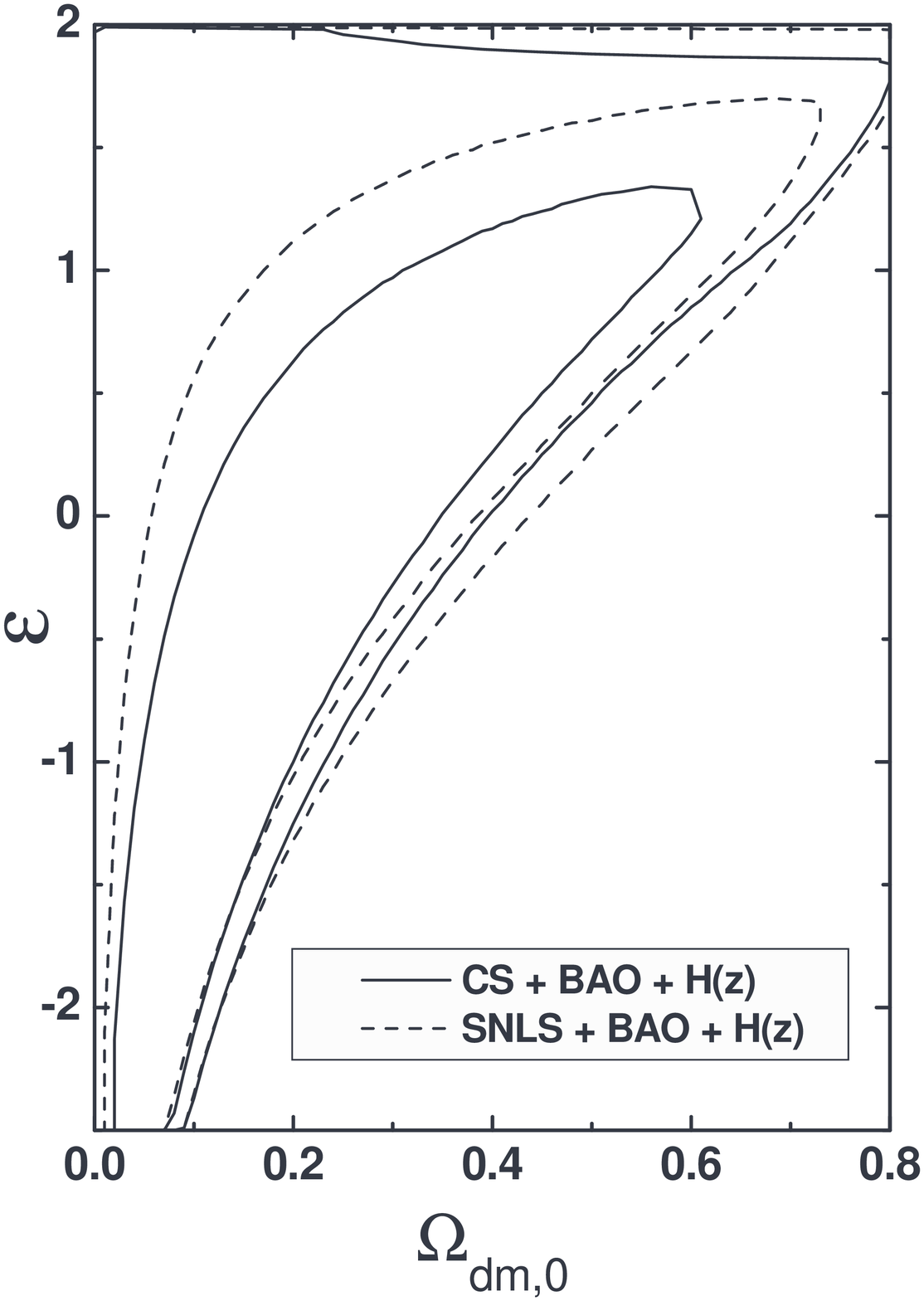,width=2.4truein,height=2.0truein}
\psfig{figure=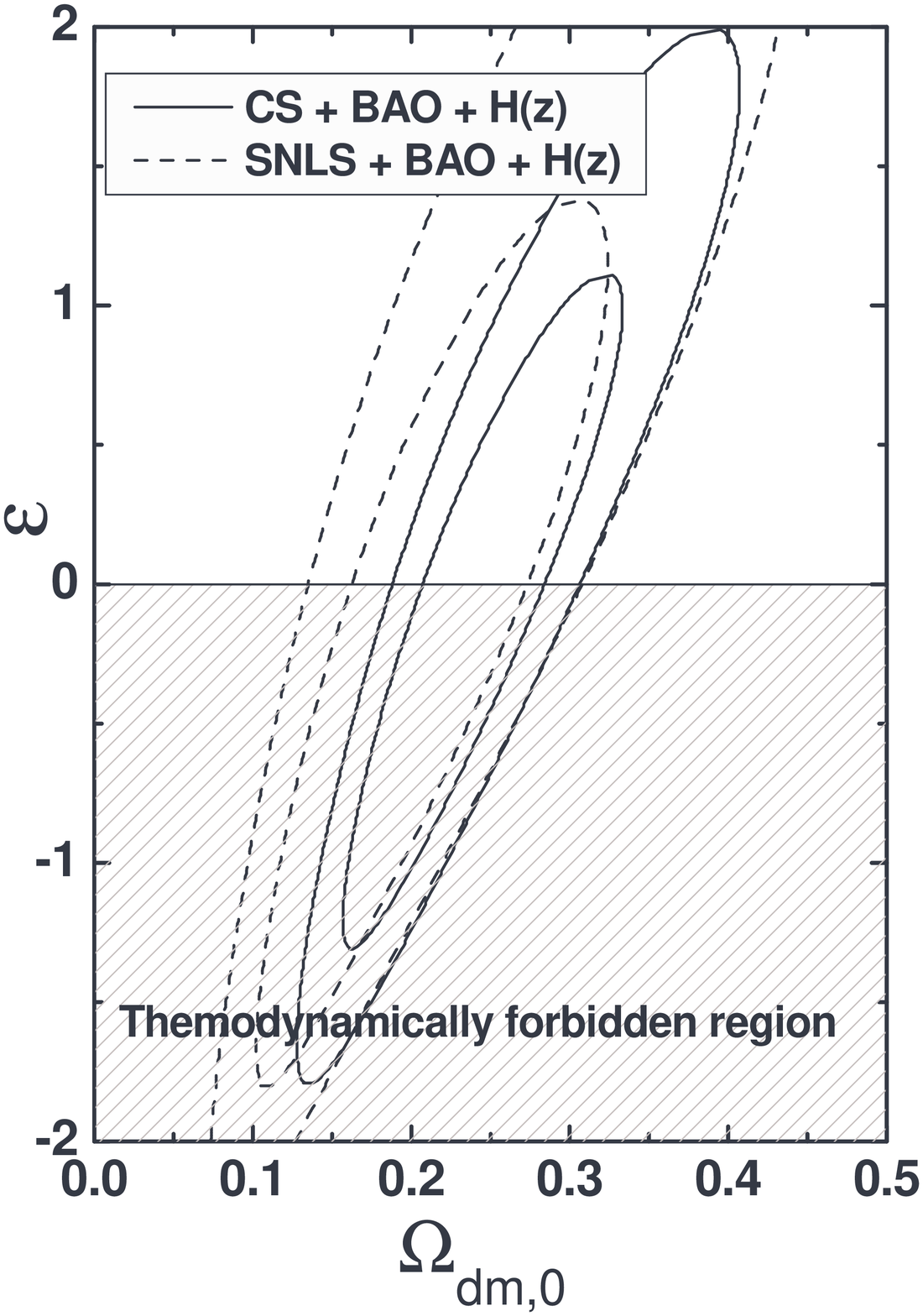,width=2.4truein,height=2.0truein} 
\hskip 0.1in} 
\caption{Contours of $\chi^2$ in the planes $w - \epsilon$ (left), $\Omega_{dm,0} - \epsilon$ (middle) and $\Omega_{dm,0} - \epsilon$ with $w = -1$ (right). These contours are drawn for $\Delta \chi^2 = 2.30$ and $6.17$. In all Panels, dashed lines correspond to  the joint analysis involving SNLS + BAO + H($z$) measurements whereas the solid ones to CS + BAO + H($z$) data. The shadowed area in the Panel at right stands for the thermodynamical constraint on $\epsilon$ discussed in  Ref.~\cite{alc}.}
\end{figure*}

By defining a new parameter
%\begin{equation}
$x \equiv {\dot\phi^2}/{(\dot\phi^2 + \rho_m)}$ with $0 \leq x \leq 1$,
%\end{equation}
we can manipulate the above equations to obtain (see \cite{jmaia} for more details)
\begin{subequations}
%\begin{equation}
%\rho_m = {3H^2 \over 8\pi G}\gamma_* (1-x)\;,
%\end{equation}
\begin{equation}\label{eq:dotphi}
\dot{\phi}^2 =  {3H^2 \over 8\pi G} \gamma_* x\;,
\end{equation}
and
\begin{equation}\label{eq:V}
V(\phi) = {3H^2 \over 8\pi G} \left[1 - \gamma_* \left(1-{x\over
2}\right)\right]\;,
\end{equation}
\end{subequations}
which link directly the field and its potential with the related quantities of the coupled quintesence case. From Eq. (\ref{eq:dotphi}), one can show that in terms of the scale factor $a$ the field $\phi$ is given by
\begin{eqnarray} \label{ph}
\phi & & = \sqrt{\frac{3}{8\pi G}}\int{\sqrt{\gamma_*(a) x}\frac{da}{a}}\;,
\end{eqnarray}
where $\gamma_*(a)$ can be obtained from Eqs. (\ref{dm}) and (\ref{de}), i.e.,
%\begin{widetext}
\begin{equation}
\label{gamma}
\gamma_* = \frac{\Omega_{b,0}a^{-3} + (1+w)\Omega_{de,0}a^{-3(1+w)} + {\rm{A}}\Omega_{dm,0}a^{-3+\epsilon}}{\Omega_{b,0}a^{-3} + \Omega_{de,0}a^{-3(1+w)} + {\rm{B}}\Omega_{dm,0}a^{-3+\epsilon}},
\end{equation}
%\end{widetext}
with ${\rm{A}} = \frac{3|w|+w\epsilon}{3|w|-\epsilon}$ and ${\rm{B}} = \frac{3|w|}{3|w|-\epsilon}$. From now on, we consider $x$ to be a constant, which is equivalent to impose the condition that the scalar field version mimics exactly the interaction rate of its phenomenological coupled counterpart \cite{jmaia}. From Eqs. (\ref{eq:V}) and (\ref{gamma}), we also obtain
\begin{eqnarray} \label{va}
U(a) & = & \frac{x}{2}\Omega_{b,0}a^{-3} + \left[\frac{x}{2}(1+w) - w \right]\Omega_{de,0}a^{-3(1+w)} \nonumber \\ & & + \left[\frac{x}{2}{\rm{A}} - w{\rm{B}}\right]\Omega_{dm,0}a^{-3+\epsilon}\;,
\end{eqnarray}
where $U = 8\pi G V(a)/3H_0^2$ and $\Omega_{de,0} = 1 - \Omega_{b,0} - {3|w|\Omega_{dm,0}/(3|w|-\epsilon)}$.

Figure (1b) shows the potential $U(\phi)$ obtained from a numerical combination of Eqs. (\ref{ph})-(\ref{va}) by assuming $\epsilon = 0.1$ and $\Omega_{dm,0} = 0.23$. Four different cases are shown, namely, quintessence ($w = -0.8$) and phantom ($w = -1.2$) EoS with and without a separately conserved baryonic component $\Omega_{b,0}$. As one may also check, the potential shown in Fig. (1) reduces to
\begin{equation} \label{vphi}
V( \phi )  = {\rho}_{de,0} + {{C_1}}\left[e^{\frac{2\phi}{{{C_2}}}} + {C_3}^2e^{-\frac{2\phi}{{\rm{C_2}}}} - 2{{C_3}} \right]\;,
\end{equation}
when $w = -1$, where $C_1$, $C_2$ and $C_3$ are constants~\cite{ernandes}. In general, we can see that the effect of $\simeq 4.16\%$ of baryons on the potential $U(\phi)$, as well as a changing in the EoS values, produces only a small shift relative to the case in which the baryonic content participates of the interacting process or to the case of vacuum decay, so that the general potential for these scenarios belongs to the same class of potentials given analytically by Eq. (\ref{vphi})~\cite{ref}.

\section{Current observational constraints}

The description of the interacting scenario discussed in Sections II and III clearly shows that it comprises a multitude of cosmological solutions for different combinations of $\epsilon$, $w$ and $\Omega_{dm,0}$. In this Section we investigate observational bounds on the parametric spaces $\epsilon-w$ and $\epsilon-\Omega_{dm,0}$ from statistical analyses involving three classes of cosmological observations. In our analysis we fix $\Omega_b=0.0416$ from WMAP results~\cite{Sperg07}, a value in good agreement with the constraints derived from primordial nucleosynthesis \cite{nucleo}.

We use current distance measurements to SNe Ia, namely, the SNLS collaboration sample of 115 points \cite{Astier06} and the most recent SNe Ia compilation, the so-called {Constitution set} (CS)~\cite{cs}. The SNLS sample includes 71 high-$z$ SNe Ia in the redshift range $0.2 < z < 1$ and 44 low-$z$ events compiled from the literature but analyzed from the same manner as the high-$z$ sample.  The CS of 397 SNe Ia covers a redshift range from $z = 0.015$ to $z = 1.551$, including 139 SNe Ia at $z < 0.08$, and constitutes the largest SNe Ia luminosity distance sample currently available.

The BAO distance ratio from 2dFGRS and SDSS measurements is given by $D_V(0.35)/D_V(0.2) = 1.812 \pm 0.060$~\cite{2df}, where~\cite{ref1}
\begin{equation}
D_V (z_{BAO}) = \left[ \frac{z_{BAO}}{H(z_{BAO})} \left(\int_{0}^{z_{BAO}}{\frac{dz}{H(z)}}\right)^2\right]^{1/3}\;,
\end{equation}
and $H(z)$ is given by Eqs. (\ref{dm}), (\ref{de}) and (\ref{fried}). In our analysis, we also use this ratio along with 9 determinations of the Hubble evolution $H(z)$, as given in Ref.~\cite{svj} (for more details on statistical analysis involving these data sets we refer the reader to Ref.~\cite{sa}). 

The results of our statistical analyses are shown in Fig. (2). As usual, the total likelihood is written as ${\cal L}\propto e^{-\chi^2/2}$, where the $\chi^2$ function takes into account the three data sets discussed above. By marginalizing ${\cal L}$ over $\Omega_{dm,0}$ ($\omega$), we can quantify how much the plane $\epsilon - \omega$ ($\epsilon - \Omega_m$) can be constrained by the data. The contour levels for these analyses are displayed, respectively, in Panels (2a) and (2b). In both Panels, dashed lines correspond to 1$\sigma$ and 2$\sigma$ values of the joint analysis involving SNLS + BAO + H($z$) measurements whereas the solid ones to CS + BAO + H($z$) data. Although the bounds on $\epsilon$ from the latter analysis (CS) are tighter than those from the former one (SNLS), the important point worth emphasizing here is that both negative and positive values for the interacting parameter are allowed by these analyses. Physically, this amounts to saying that not only an energy flow from dark energy to dark matter ($\epsilon > 0$) is observationally allowed but also a flow from dark matter to dark energy ($\epsilon < 0$) [see Eq. (\ref{dm})]. At 95.4\% (C.L.), we have found $w = -0.92 \pm 0.15$ and $\epsilon =-0.84_{-0.71}^{+1.07}$ and $w = -0.78 \pm 0.08$ and $\epsilon = -1.54_{-0.88}^{+0.97}$, respectively, for the SNLS + BAO + H($z$) and CS + BAO + H($z$) joint analyses.

In Fig. (2c) we show a similar analysis by fixing the dark energy EoS at $w = -1$, which is fully equivalent to the vacuum decay scenario discussed in Refs.~\cite{wm,alc}. As shown in Ref.~\cite{alc}, for the vacuum decay process the interacting parameter $\epsilon$ is restricted by thermodynamical arguments to be positive~\cite{ref2}. By considering this physical constraint on $\epsilon$ we have found at 95.4\% (C.L.) $\epsilon = 0.0_{-0.00}^{+0.89}$ and $\Omega_{dm,0} = 0.21 \pm 0.06$ [SNLS + BAO + H($z$)] and $\epsilon = 0.0^{+0.58}_{-0.00}$ and $\Omega_{dm,0} = 0.24 \pm 0.04$  [CS + BAO + H($z$)], which is equivalent to the standard $\Lambda$CDM case. If we drop the thermodynamical constraint ($\epsilon > 0$), we find, respectively, $\epsilon =-0.46_{-0.68}^{+0.91}$ and $\Omega_{dm,0} = 0.19 \pm 0.05$ and $\epsilon = -0.24_{-0.53}^{+0.60}$ and  $\Omega_{dm,0} = 0.23 \pm 0.04$ at 95.4\% (C.L.).

\section{Final Remarks}

We have discussed some cosmological consequences of an alternative mechanism of cosmic acceleration based on a general class of coupled quintessence scenarios whose interaction term is deduced from the effect of the dark energy on the CDM expansion rate. The resulting expressions for the model are parameterized by a small parameter $\epsilon$ and have many of the previous phenomenological approaches as a particular case. 

Following the procedure proposed in Refs.~\cite{jmaia,jmaia1}, we have also discussed a scalar field description for these cosmologies and found that this class of models is identified with a coupled quintessence field with the potential given by Eqs. (\ref{ph})-(\ref{va}) [see also Fig. (1)]. From the observational point of view, we have investigated the current bounds on the interacting parameter $\epsilon$ from the most recent data of SNe Ia (SNLS and Constitution samples), baryon acoustic oscillation peak at $z = 0.20$ (2dFGRS) and $z = 0.35$ (SDSS), and the Hubble evolution $H(z)$. We have shown that negative and positive values for the interacting parameter $\epsilon$ are observationally allowed, which means that both an energy flow from dark energy to dark matter as well as a flow from dark matter to dark energy are possible. It is worth emphasizing that in our analysis the interacting parameter has been considered constant whereas in a more realistic case it must vary with redshift. The theoretical and observational consequences of this more realistic interacting scenario, as well as a full comparison with the case discussed in the present analysis, will appear in a forthcoming communication~\cite{new}.

\acknowledgments
FEMC is grateful to CAPES - Brazil for financial support. EMB and JSA thank CNPq - Brazil for the grants under which this work was carried out.


\begin{thebibliography}{99}
\bibitem{ref} Double exponential potentials of the type (\ref{vphi}) have been considered in the literature as viable examples of quintessence scenarios (see, e.g, \cite{dp}). As discussed in Ref.~\cite{dp1}, a field potential given by the sum of two exponential terms is also motivated by dimensional reduction in M-theory with implications for the late-time accelerating behavior of the cosmic expansion.
\bibitem{ref1} As discussed in Ref.~\cite{doran}, the BAO parameter ${\cal{A}} = D_V\sqrt{\Omega_mH_0^2}/zc$ can be considered as a good approximation for a large class of dark energy models. In our analysis, we do not use this quantity because it also uses the approximation that the sound travel distance is $\propto 1/\sqrt{\Omega_{m}}$, which fails if dark energy has a significant contribution in the past. According to Fig. (1a), in our interacting scenarios it may happen for relatively large and positive values of $\epsilon$ (see \cite{saulo} for a discussion).
\bibitem{ref2} Note that the thermodynamical constraint $\epsilon > 0$ seems not to apply to a general quintessence case ($w \neq 1$). See, e.g.,  Refs.~\cite{t1} and \cite{t2} for a discussion.
\end{thebibliography}
\end{document}